\documentclass[amstex,twocolumn,showpacs,floats,floatfix,aps,pra]{revtex4}
\usepackage{amssymb}
\usepackage{amsmath}
\usepackage{calc}
\usepackage{graphicx}
\usepackage{bm}
\usepackage{amsfonts}
\usepackage{color}
\usepackage{epsfig}

\begin{document}

\author{A. A. Rangelov}
\affiliation{Department of Physics, Sofia University, James Bourchier 5 Blvd, 1164
Sofia, Bulgaria}
\title{Magnetic levitation induced by negative permeability}

\begin{abstract}
In this paper we study the interaction between a point magnetic
dipole and a semi-infinite metamaterial using the method of
images. We obtain analytical expressions for the levitation force
for an arbitrarily oriented dipole. Surprisingly the maximal
levitation force for negative permeability is found to be stronger
compared to the case when the dipole is above a superconductor.
\end{abstract}

\maketitle


Naturally materials have positive permeability, but for the last decade
artificial materials with negative permeability (metamaterials) attract
growing interest from theoretical and experimental perspectives \cite%
{Pendry1,Wiltshire}. In such materials the permeability $\mu /\mu _{0}$ ($%
\mu _{0}$-vacuum permeability) is given by%
\begin{equation}
\mu /\mu _{0}=1-\alpha \omega _{0}^{2}/\left( \omega ^{2}-\omega
_{0}^{2}\right) ,  \label{omega}
\end{equation}%
where $\omega _{0}$ is analog to the plasma frequency (called
\textquotedblleft magnetic plasma frequency\textquotedblright ) and $\alpha $
is the so-called filling factor \cite{Pendry1,Wiltshire}. It is not hard to
check that for some frequencies $\omega $ the permeability $\mu $ is negative.

Most of the papers, related to negative permeability, focus on the
optical properties of those metamaterials \cite{Pendry2,Klein,Wee}. In the
present paper we consider the magnetic forces in the presence of material
with negative permeability. We explore theoretically the force acting on a
point magnetic dipole placed next to a material with negative permeability.
Our examination relies on the Pendry's criterion for validity of
magnetostatic limit in the present problem \cite{Pendry3}: the wavelength,
corresponding to the frequency $\omega $ in Eq. (\ref{omega}), should be
longer than the distance between the magnetic dipole and the material with
negative permeability. Using the method of images \cite%
{Griffiths,Jackson,Knoepfel}, we show that for negative
permeability materials the force between the metamaterial and the
magnetic dipole can be attractive or repulsive and can even have a
much bigger value compared to the conventional materials.

\begin{figure}[tb]
\centerline{\epsfig{width=75mm,file=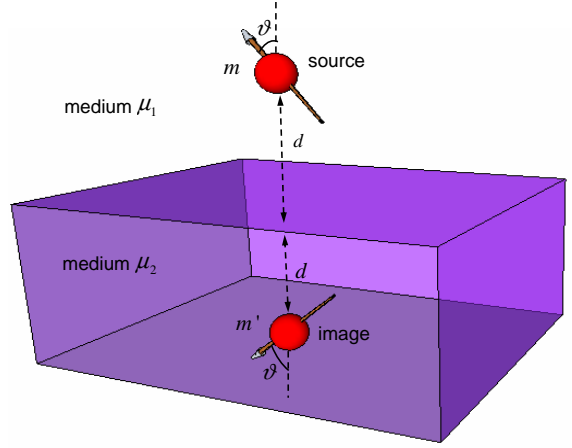}}
\caption{Magnetic dipole $m$ embedded in a semi-infinite material with
permeability $\protect\mu _{1}$ at a distance $d$ away from a plane
interface that separates the two mediums and its dipole image $m^{\prime }$
symmetrically placed in a semi-infinite material with permeability $\protect%
\mu _{2}$.} \label{method of images}
\end{figure}

We consider a point magnetic dipole $m$ in media with permeability $\mu _{1}$%
, oriented at angle $\theta $ with respect to the axis perpendicular to the
plane surface of a medium with permeability $\mu _{2}$ as shown in Fig. \ref%
{method of images}. The distance from the dipole to the surface,
that separate the two mediums, is $d$ (see Fig. \ref{method of
images}). The problem we set is to find how the dipole $m$
interacts with the medium $\mu _{2}$. The
solution is quite straightforward when one uses the method of images \cite%
{Griffiths,Jackson,Knoepfel}, which states that the magnetic potential in
the $\mu _{1}$ region is a superposition of the dipole fields from $m$ and
the image $m^{\prime }$%
\begin{equation}
m^{\prime }=\frac{\mu _{1}-\mu _{2}}{\mu _{1}+\mu _{2}}m,
\end{equation}%
placed at the same distance\ $d$ on the other side of the plane separating
the two mediums and oriented at the same angle $\theta $ with the
perpendicular axis to the plane surface (see Fig. \ref{method of images}).

Using the well-known expression for the potential energy of two real
magnetic dipoles $m$ and $m^{\prime }$ at a distance $\mathbf{r}$ \cite%
{Yang,Coffey}
\begin{equation}
V=\frac{3\mu _{1}}{4\pi }\left[ \frac{\mathbf{m.m}^{\prime }}{\left\vert
r\right\vert ^{3}}-\frac{3\left( \mathbf{r\cdot m}\right) \left( \mathbf{r.m}%
^{\prime }\right) }{\left\vert r\right\vert ^{5}}\right] ,
\end{equation}%
we obtain the following expression for the vertical force acting on a dipole $m$
\begin{equation}
F=\frac{3\mu _{1}}{4\pi }\frac{\left( 1+\cos ^{2}\theta \right) mm^{\prime }%
}{\left( 2d\right) ^{4}}=\frac{3\mu _{1}m^{2}}{4\pi }\frac{\left( 1+\cos
^{2}\theta \right) }{\left( 2d\right) ^{4}}\frac{\mu _{1}-\mu _{2}}{\mu
_{1}+\mu _{2}},  \label{net magnetical force}
\end{equation}%
which force is perpendicular to the surface of a medium with permeability $%
\mu _{2}$ and is attractive or repulsive depending on the ratio $\left(
\mu _{1}^{2}-\mu _{1}\mu _{2}\right) /\left( \mu _{1}+\mu _{2}\right) $. The
repulsion force for conventional materials is strongest when the magnetic
dipole $m$ is in free space, pointing towards the surface plane and is placed
next to a superconductor. This is the Meissner's effect \cite%
{Knoepfel,Kordyuk,Lin,Perez}, which follows formally from Eq. (\ref{net
magnetical force}) by setting $\theta =0$, $\mu _{1}=\mu _{0}$ and $\mu
_{2}=0$ in this case the vertical lifting force is%
\begin{equation}
f=\frac{3\mu _{0}m^{2}}{2\pi \left( 2d\right) ^{4}}.  \label{Meissner effect}
\end{equation}

If we consider the case when the first media is vacuum ($\mu _{1}=\mu _{0}$)
while the second media is material with permeability $\mu _{2}=\mu $
(material or metamaterial), the force acting on a magnetic dipole is%
\begin{equation}
F=\frac{3\mu _{0}m^{2}}{4\pi }\frac{\left( 1+\cos ^{2}\theta \right) }{%
\left( 2d\right) ^{4}}\frac{\mu _{0}-\mu }{\mu _{0}+\mu }.
\label{net metamaterial force}
\end{equation}%
This force is repulsive when%
\begin{equation}
\left\vert \mu \right\vert <\mu _{0}
\end{equation}%
and attractive when%
\begin{equation}
\left\vert \mu \right\vert >\mu _{0}.
\end{equation}%
Eq. (\ref{net metamaterial force}) has singularity for $\mu=-\mu _{0}$.
One may think that it leads to an infinite force,
but one should note that dispersionless material is an abstraction and a finite
positive imaginary component of the permeability always exists \cite{Pendry3}%
. The imaginary component of permeability can no longer be neglected when
the difference between the real permeabilities is small, which
resolves the infinite force paradox.

\begin{figure}[tb]
\centerline{\epsfig{width=75mm,file=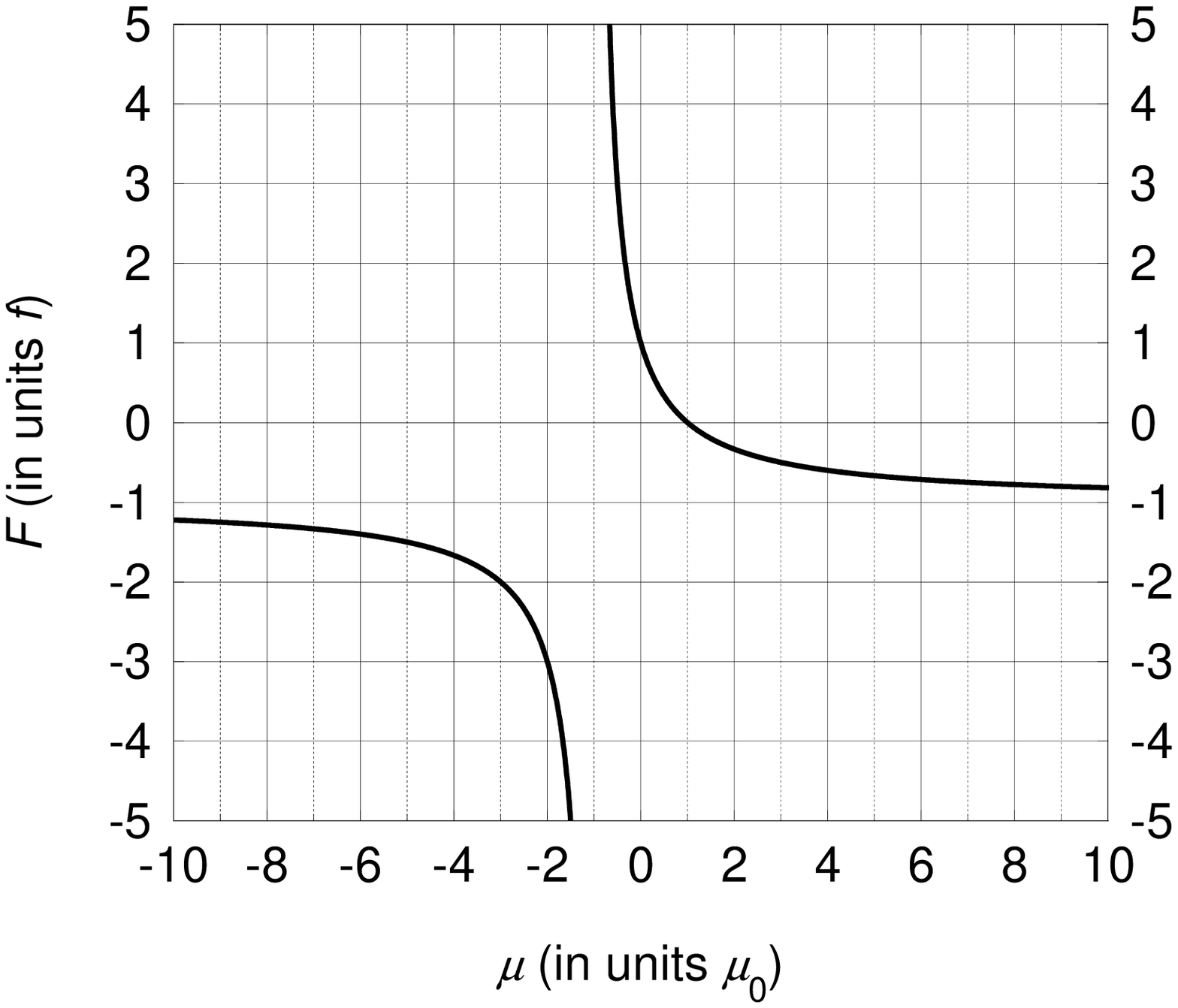}} \caption{Maximal
vertical force, the case $\protect\theta =0$ from Eq. (\protect
\ref{net metamaterial force}), as a function of the permeability
$\protect\mu $. The force is divided by the maximal force $f$,
which is due to Meissner's effect from Eq. (\protect\ref{Meissner
effect}).} \label{fig2}
\end{figure}

Even more interesting than the direction of the force is that the force
could be stronger than the one due to the Meissner effect, which can be
seen from Fig. \ref{fig2}. For example, when $\mu =-2\mu _{0}$ the
dipole is attracted to the metamaterial with maximal force
\begin{equation}
F=-\frac{9\mu _{0}m^{2}}{2\pi \left( 2d\right) ^{4}}=-3f.
\end{equation}%
The last equation and Fig. \ref{fig2} makes it clear that the
levitation force acting on a magnetic dipole can be much stronger
than the levitation force on the same dipole above a
superconductor. Moreover the levitation in case of metamaterials
can be done at room temperature, in contrast to the superconductor
levitation case.

The examined force of a point magnetic dipole above a negative permeability
has the potential to be not only a curious and intriguing example of what
artificial materials can do, but also a useful and effective technique to
levitate small objects and therefore to make frictionless devices.

This work has been supported by the European Commission projects EMALI and
FASTQUAST, the Bulgarian NSF grants D002-90/08 and DMU02-19/09.

\end{document}